\documentclass[]{aa}
\usepackage{graphicx}
\usepackage{txfonts}

\begin{document}

\title{Two-wave interaction in ideal magnetohydrodynamics}

\author{
T.V. Zaqarashvili\inst{1} \and B. Roberts\inst{2}}

\institute{Georgian National Astrophysical Observatory (Abastumani
Astrophysical Observatory), Kazbegi Ave. 2a, Tbilisi 0160, Georgia; \\
\email{temury@genao.org} \and School of Mathematics and Statistics,
University of St. Andrews, St. Andrews, Fife, KY16 9SS, Scotland, U
K; \email{bernie@mcs.st-and.ac.uk}}

\offprints{T.V. Zaqarashvili}

\date{Received / Accepted }

\abstract{The weakly nonlinear interaction of sound and linearly
polarised Alfv{\'e}n waves propagating in the same direction along
an applied magnetic field is studied. It is found that a sound wave
is coupled to the Alfv{\'e}n wave with double period and wavelength
when the sound and Alfv{\'e}n speeds are equal. The Alfv{\'e}n wave
drives the sound wave through the ponderomotive force, while the
sound wave returns energy back to the Alfv{\'e}n wave through the
parametric (swing) influence. As a result, the two waves alternately
exchange their energy during propagation. The process of energy
exchange is faster for waves with stronger amplitudes. The
phenomenon can be of importance in astrophysical plasmas, including
the solar atmosphere and solar wind.
}

\maketitle

\section{Introduction}

Magnetohydrodynamic (MHD) waves play an important role in various
astrophysical situations. The linear description of small amplitude
waves is often a good approximation. However, highly dynamical and
inhomogeneous astrophysical plasmas may permit the interaction
between different MHD wave modes, leading to basically new
processes. The energy transformation from one kind of wave into
another is of importance in the processes of energy transport and
dissipation. MHD wave coupling due to the medium inhomogeneity has
been extensively studied in solar-terrestrial connections (Chen \&
Hasegawa \cite{che}; Ionson \cite{ion}; Rae \& Roberts \cite{rae};
Heyvaerts \& Priest \cite{hey}; Hollweg \cite{hol1}; Poedts et al.
\cite{poedts}; Ofman \& Davila \cite{ofm}). It is also known that in
a homogeneous plasma large amplitude circularly polarised Alfv{\'e}n
waves may resonantly drive two other waves, through the three-wave
interaction (or decay instability) (Galeev and Oraevsky \cite{gal};
Sagdeev \& Galeev \cite{sag}). The interaction between various kinds
of magnetosonic and Alfv{\'e}n waves in a uniform plasma has been
studied previously (Brodin \& Stenflo \cite{bro}). On the other
hand, Alfv{\'e}n waves with varying magnetic field strength (e.g.,
linearly and elliptically polarised waves) may directly drive
density fluctuations through the ponderomotive force (Hollweg
\cite{hol}; Cohen \& Kulsrud \cite{coh}). When the Alfv{\'e}n speed
$v_A$ is different from the sound speed $v_s$, then the frequency
and spatial inhomogeneity of generated density fluctuations do not
satisfy the dispersion relation of sound waves, and therefore cannot
propagate as sound waves, but instead they modify the Alfv{\'e}n
speed. However, when $v_A \approx v_s$, then the density
perturbations may propagate as sound waves with a frequency and wave
number that is double that of the Alfv{\'e}n waves. As a
result, the nonlinear Schr{\"o}dinger equation describing the
propagation of finite amplitude Alfv{\'e}n waves along the magnetic
field exhibits a discontinuity in this particular case (Spangler \&
Sheerin \cite{spa}; Medvedev et al. \cite{med}).
Medvedev et al. (\cite{med}) numerically solved the kinetic nonlinear
Schr{\"o}dinger equation in this case and showed the damping of
coherent Alfv{\'e}n wave trains. However, the back-reaction of
generated sound waves (or ion-acoustic waves) on an Alfv{\'e}n wave is
not well studied. We can expect that a nonlinear
coupling between Alfv{\'e}n and sound waves may take place in this
particular case, but no clear physical description of the process
has so far been given.

On the other hand, the recently proposed new mechanism of MHD swing
(wave-wave) interaction (Zaqarashvili \cite{zaq1}; Zaqarashvili \&
Roberts \cite{zaq2}; Shergelashvili et al. \cite{she}) can be
responsible for the energy transformation from magnetosonic waves
into Alfv{\'e}n waves. The physical basis of this interaction is the
parametric influence: the magnetosonic waves cause a periodical
variation in the medium parameters, which in turn influences the
phase velocity of transversal Alfv{\'e}n waves and this leads to a
resonant energy transformation into certain harmonics. It has been
shown that a periodical variation of the medium density, caused by
the propagation of sound waves along an applied magnetic field in a
high $\beta$ plasma ($\beta={8{\pi}p}/{B^2}$, where $p$ is the
plasma pressure and $B$ is the magnetic field), results in
Alfv{\'e}n waves governed by the Mathieu equation (Zaqarashvili
\cite{zaq1}). Consequently, harmonics with half the frequency of
sound waves grow exponentially in time. The phenomenon has been
developed for the case of fast magnetosonic waves propagating across
a magnetic field and Alfv{\'e}n waves propagating along the field
(Zaqarashvili \& Roberts \cite{zaq2}; Shergelashvili et al.
\cite{she}). The process of energy exchange between these different
kinds of wave motion is called {\it swing wave-wave interaction}
(Zaqarashvili \& Roberts \cite{zaq2}). This terminology arises by
analogy with a spring pendulum. It can be shown that the
transversal (pendulum) and longitudinal (spring) oscillations are
coupled when the eigenfrequency of transversal oscillations is half
the frequency of the spring oscillations.

Taking into account the swing wave-wave interaction, one may suggest
that sound waves resonantly drive the harmonics of Alfv{\'e}n waves
with half frequency and wave number when $v_A \approx v_s$.
Consequently, the waves may alternately exchange energy as they
propagate along an applied magnetic field. A sound wave entering a
region where $v_A \approx v_s$ may resonantly drive an Alfv{\'e}n
wave, and {\it vice versa}. Also, the plasma $\beta$ approaches to
unity in parts of the solar wind plasma, and in the lower solar
atmosphere there is a region where the 
Alfv{\'e}n and sound speeds are approximately
equal (Gary \cite{gary}; see Fig. 3 of that paper), Moreover, the recent
observations by Muglach et al. (\cite{mug}) suggest a possible
transformation of compressible wave energy into incompressible waves
in this particular region. Therefore the process of Alfv{\'e}n and
sound wave coupling at $v_A \approx v_s$ can be of general importance in the
solar atmosphere, solar wind and other astrophysical situations.

In the next section we consider analytically the weakly nonlinear MHD
equations, giving a short description of some astrophysical
situations where the process can be of importance.

\section{Wave Interactions}

Consider fluid motions ${\bf u}$ in a magnetised medium (with zero
viscosity and infinite electrical conductivity), as described by the
ideal MHD equations:
\begin{equation}
{{{\partial \bf B}}\over {\partial t}} + ({\bf
u}{\cdot}{\nabla}){\bf B}= ({\bf B}{\cdot}{\nabla}){\bf u} - {\bf
B}{\nabla}{\cdot}{\bf u},\,\,\,\,\,{\nabla}{\cdot}{\bf B}=0,
\end{equation}
\begin{equation}
{\rho}{{{\partial \bf u}}\over {\partial t}} + {\rho}({\bf
u}{\cdot}{\nabla}) {\bf u} = - {\bf {\nabla}}\left[p + {{B^2}\over
{8{\pi}}}\right ] + {{({{\bf B}{\cdot}{\nabla}}){\bf B}}\over
{4{\pi}}},
\end{equation}
\begin{equation}
{{{\partial {\rho}}}\over {\partial t}} + ({\bf
u}{\cdot}{\nabla}){\rho}
 + {\rho}{\nabla}{\cdot}{\bf u}=0,
\end{equation}
\begin{equation}
{{{\partial p}}\over {\partial t}} + ({\bf u}{\cdot}{\nabla}){p}  +
\gamma p{\nabla}{\cdot}{\bf u}=0,
\end{equation}
where $p$ is the pressure, $\rho$ is the density, ${\bf B}$ is the
magnetic field and $\gamma$ is the ratio of specific heats. We
neglect gravity, though it may be of importance in some
astrophysical situations.

Consider the case of a homogeneous medium with an uniform magnetic
field ${\bf B}=(B_0,0,0)$ directed along the $x$ axis of a Cartesian
coordinate system.
We consider wave propagation along the $x$ axes (thus along the
magnetic field) and wave polarisation in $xy$ plane. In this case,
two kinds of wave may propagate strictly along the applied magnetic
field: sound and linearly polarised Alfv{\'e}n waves. In the linear
limit these waves are strictly different; the Alfv{\'e}n waves are
purely transversal with the velocity component along the $y$ axis
(magnetic tension provides the restoring force), while the sound
waves are purely longitudinal with the velocity component along the
$x$ axis (pressure gradients provide the restoring force). In this
case the energy exchange between waves results from nonlinear
interactions. As velocity and magnetic field components of linearly
polarised Alfv{\'e}n waves lay in $xy$ plane, then we may consider
the two dimensional MHD equations.

In principle, the transverse inhomogeneity of the magnetic field
along the $z$-axis can also be taken into account. In this case each
magnetic $xy$ surface will behave independently (when the waves
propagate strictly along the applied field and the dissipation is
neglected). Therefore we may again consider two dimensional
equations for each magnetic $xy$ surface separately. However, 
here we consider the homogeneous medium, keeping in mind
that similar processes can also occur in inhomogeneous
(transverse to the magnetic field) plasmas.

As we consider the wave propagation along the $x$ axis, then $x$ and
$y$ components of equations (1)-(4) take the form:
\begin{equation}
{{{\partial b_y}}\over {\partial t}} + u_x{{{\partial b_y}}\over
{\partial x}} + b_y{{{\partial u_x}}\over {\partial x}} -
B_0{{{\partial u_y}}\over {\partial x}}=0,
\end{equation}
\begin{equation}
{\rho}{{{\partial u_y}}\over {\partial t}} + {\rho}u_x{{{\partial
{u_y}}}\over {\partial x}} - {{B_0}\over {4\pi}}{{\partial b_y}\over
{\partial x}}=0,
\end{equation}
\begin{equation}
{{{\partial {\rho}}}\over {\partial t}} + {\rho}{{{\partial
{u_x}}}\over {\partial x}} + u_x{{{\partial {\rho}}}\over {\partial
x}}=0,
\end{equation}
\begin{equation}
{\rho}{{{\partial u_x}}\over {\partial t}} + {\rho}u_x{{{\partial
{u_x}}}\over {\partial x}} + {{{\partial p}}\over {\partial x}} +
{{\partial}\over {\partial x}}{{b^2_y}\over {8\pi}}=0,
\end{equation}
\begin{equation}
{{{\partial p}}\over {\partial t}} + {\gamma}p{{{\partial
{u_x}}}\over {\partial x}} + u_x{{{\partial p}}\over {\partial
x}}=0,
\end{equation}
where $p$ and $\rho$ denote the total pressure and density, $u_y$
and $u_x$ are the velocity perturbations (of the Alfv{\'e}n and
sound waves, respectively), and $b_y$ is the perturbation in the
magnetic field. These equations describe the fully nonlinear
behaviour of sound and linearly polarised Alfv{\'e}n waves
propagating along an applied magnetic field.

For a mechanical analogy of the wave coupling process, we recall the
pendulum with stiffness spring (Zaqarashvili \& Roberts
\cite{zaq2}). There are two different oscillations of such a
pendulum: transversal oscillations are due to gravity, and
longitudinal oscillations are due to the stiffness force of the
elastic spring. Transversal oscillations influence the longitudinal
ones, due to the varying gravitational field component acting along
the pendulum axis, while longitudinal oscillations influence
transversal ones parametrically through the variation of the
pendulum length. When the eigenfrequency of the transversal
oscillation is half the frequency of the spring oscillation, then a
resonant coupling occurs. Initial transversal oscillations directly
drive longitudinal oscillations and they return the energy back to
transversal oscillations through the parametric influence. The
energy exchange between the oscillations occurs alternately, without
dissipation.

\begin{figure}
\includegraphics[width=7cm]{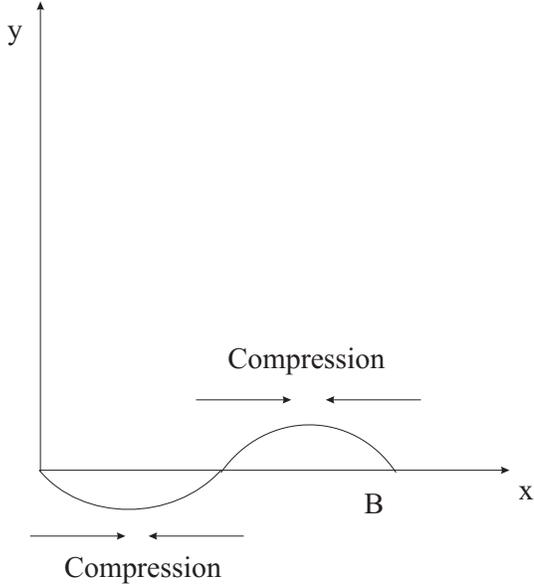}
\caption{\label{fig:epsart}The influence of a sound wave on an
Alfv{\'e}n wave with double period and wavelength. The periodical
compression (expansion) of the medium due to the sound wave may
amplify the Alfv{\'e}n wave. The double period of the Alfv{\'e}n
wave arises due to the displacement of plasma elements on both sides
of the magnetic field.}
\end{figure}

A similar process occurs in the case of Alfv{\'e}n and sound waves
when they propagate with the same phase speeds along an applied
magnetic field. If initially there is an Alfv{\'e}n wave, then it
drives sound waves with twice the frequency and wave number of the
Alfv{\'e}n wave through the ponderomotive force (Hollweg
\cite{hol}). A more interesting and new process occurs when
initially there is a sound wave: then small transversal
perturbations can be amplified through the parametric action of the
sound wave (Zaqarashvili \cite{zaq1}). The Alfv{\'e}n wave results
from the Lorentz force acting against the fluid inertia. The
periodical compression (expansion) of the medium due to the sound
wave reinforces the fluid inertia in certain phases and this leads
to the amplification of an Alfv{\'e}n wave with half the frequency
and wave number of the sound wave (see Figure 1). The parametric
influence of sound waves is expressed through the nonlinear terms in
equations (5) and (6).

We express the total plasma density and pressure as the sum of the
unperturbed and perturbed parts, $\rho_0 + \rho_1$ and $p_0 + p_1$,
respectively, and consider the weakly nonlinear process when the
perturbations are much smaller than the unperturbed values. Then the
variation of wave amplitudes, due to nonlinear interactions, will be
a slow process. The perturbations can be represented as the product
of a slowly varying amplitude $C_j(t)$ (with complex conjugate
$C^{*}_j(t)$) and a rapidly oscillating term:
\begin{eqnarray}
\rho_1=C_1(t)e^{i\phi_2(t,x)} +
C^{*}_1(t)e^{-i\phi_2(t,x)},\nonumber\\ u_x=C_2(t)e^{i\phi_2(t,x)}
+ C^{*}_2(t)e^{-i\phi_2(t,x)},\nonumber\\
b_y=C_3(t)e^{i\phi_1(t,x)} + C^{*}_3(t)e^{-i\phi_1(t,x)},\nonumber\\
u_y=C_4(t)e^{i\phi_1(t,x)} + C^{*}_4(t)e^{-i\phi_1(t,x)},\nonumber\\
p_1=C_5(t)e^{i\phi_2(t,x)} + C^{*}_5(t)e^{-i\phi_2(t,x)}.
\end{eqnarray}
Here
\begin{equation}
{\phi_1}={\omega_A}t - k_Ax,~~~~{\phi_2}={\omega_s}t - k_sx,
\end{equation}
 where ${\omega_A}$, $k_A$, ${\omega_s}$, $k_s$ are the
frequencies and wave numbers of Alfv{\'e}n and sound waves,
respectively. The frequencies and wave numbers satisfy the
dispersion relations for sound and Alfv{\'e}n waves:
$\omega_s/k_s=v_s$ and $\omega_A/k_A=v_A$, where $v_s=\sqrt{\gamma
p_0/\rho_0}$ and $v_A=B_0/\sqrt{4 \pi \rho_0}$ are the sound and
Alfv{\'e}n speeds respectively.

In general, the amplitudes $C_j$ can be slowly varying functions of
the $x$ coordinate too. However, for simplicity we neglect the $x$
dependence and look to their temporal evolution (see Sagdeev \&
Galeev (\cite{sag}) for a similar approximation).

Substitution of expressions (10) into equations (5)-(9), and
averaging over rapid oscillations in $x$ and $t$ (as both phases
${\phi_1}$ and ${\phi_2}$ include these variables), leads to the
cancelling of all exponential terms (after averaging, harmonic
functions become zero), so that only the first and third order (in
$C_j$) terms remain. In the first approximation (neglecting all
third order terms), the sound and Alfv{\'e}n waves are decoupled and
the amplitudes $C_j$ are constant. Third order terms with $C_j$
(which are due to the advective terms in the momentum equation) are
significant only in the case of very large amplitudes and presumably
lead to the steepening of wave fronts and consequently to the
formation of shock waves. However, if frequencies and wave numbers
satisfy the conditions
\begin{equation}
\omega_s=2\omega_A,~~~~k_s=2k_A ,
\end{equation}
or $\omega_s=-2\omega_A$ and $k_s=-2k_A$, then after averaging the
second order terms (with $C_j$) remain and wave amplitudes become
time dependent. These conditions are fulfilled when
\begin{equation}
v_A=v_s.
\end{equation}
Therefore the sound and Alfv{\'e}n waves are resonantly coupled
and so may exchange energies when they propagate with the same
speed along the magnetic field. If wave phase speeds vary across
the magnetic field, then the wave coupling may take place in a
particular {\it resonant} layer where condition (13) is satisfied.

With conditions (12)-(13), the averaging of equations (5)-(9) over
rapid oscillations in $x$ and $t$ leads to equations which govern
the temporal behaviour of complex amplitudes (with the third order
terms in $C_j$ neglected):
\begin{equation}
{{d C_3}\over dt} + i{\omega_A}C_3 + ik_AB_0C_4 - ik_AC_2C^*_3 =0,
\end{equation}
\begin{equation}{{d C_4}\over dt} + i{\omega_A}C_4 + i{B_0\over
{4\pi\rho_0}}k_AC_3 + {C_1\over \rho_0}{{d C^*_4}\over dt} -
i\omega_A{C_1\over \rho_0}C^*_4 + ik_AC_2C^*_4=0,
\end{equation}
\begin{equation}
{{d C_1}\over dt} + 2i{\omega_A}C_1 - 2ik_A\rho_0C_2 =0,
\end{equation}
\begin{equation}
{{d C_2}\over dt} + 2i{\omega_A}C_2 - 2ik_Av^2_s{C_1\over {\rho_0}}
- {{2ik_AC^2_3}\over {8\pi\rho_0}}=0.
\end{equation}
These equations describe the process of energy exchange between the
sound and Alfv{\'e}n waves. The influence of the ponderomotive force
is expressed by the last term in equation (17) and the parametric
influence of the sound wave is expressed through the last term in
equation (14) and the last three terms in equation (15).

From equations (14)-(17) we obtain two second order differential
equations by eliminating $C_2$ and $C_4$ in favour of $C_1$ and
$C_3$ (third order terms in $C_j$ being neglected):
$${{d^2 C_3}\over dt^2} + 2i{\omega_A}{{d C_3}\over dt} -
{1\over {2\rho_0}}C^*_3{{d^2 C_1}\over dt^2} - {1\over {\rho_0}} {{d
C_1}\over dt}{{d C^*_3}\over dt} - {{i{\omega_A}}\over
{\rho_0}}C^*_3{{d C_1}\over dt} + 
$$
\begin{equation}
+{{\omega^2_A}\over
{\rho_0}}C_1C^*_3 - {1\over {\rho_0}}C_1{{d^2 C^*_3}\over dt^2} = 0
\end{equation}
and
\begin{equation}
{{d^2 C_1}\over dt^2} + 4i{\omega_A}{{d C_1}\over dt} +
{{k^2_A}\over {2\pi}}C^2_3=0.
\end{equation}
Analytical solution of these equations is still complicated.
However, if the amplitude of one wave is considered to be stronger
than the other, then a significant simplification can be achieved.
In this case the back reaction can be ignored (at least at the
initial stage) and the amplitude of the initial wave becomes
constant. This allows us to study the influence of the pumping wave
on the evolution of the other wave.

We begin by studying the general behaviour of equations (18)-(19),
using a phase-plane analysis. As $C_1$ and $C_3$ are slowly varying
functions in time, we may neglect the second derivatives and third
order terms in $C_3$. Then
\begin{equation}
2i{\omega_A}{{d C_3}\over dt} + {{\omega^2_A}\over {\rho_0}}C_1C^*_3
= 0,
\end{equation}
\begin{equation}
4i{\omega_A}{{d C_1}\over dt} + {{k^2_A}\over {2\pi}}C^2_3=0.
\end{equation}

Writing $C_{1} = C_{10}+iC_{11}$ and $C_3 = C_{30}+iC_{31}$, we note
that the equations simplify if we choose the initial phase of the
Alfv{\'e}n waves so that $C_{30}=C_{31}$. Then, from equation (21),
we obtain $C_{10}=0$. Writing $A=C_{10}/\rho_0$ and $B=C_{30}/B_0$,
we have
\begin{equation}
{{d A}\over {dt}} = -{\omega_A}B^2(t),
\end{equation}
\begin{equation}
{{d B}\over {dt}} = {1\over 2}{\omega_A}A(t)B(t).
\end{equation}
These equations describe a so called autonomous system as the time
variable does not appear in the right-handside of the equations. The
general behaviour of such a system is revealed by phase-plane
analysis (see Jordan and Smith \cite{jor}, pp. 36-38). The solutions
$A(t), B(t)$ may be represented on a plane with cartesian axes $A$,
$B$. Then as $t$ increases the point $(A(t), B(t))$ traces out a
directed curve in the plane; this is a {\it phase path}. From
equations (22)-(23) we get:
\begin{equation}
{{dB}\over {dA}} = - {1\over {2}}{A\over B},
\end{equation}
which integrates to
\begin{equation}
 A^2 + 2B^2 = const.
\end{equation}
The phase diagram of the system is plotted in Fig. 2. Note that the
equilibrium points of this system (where the right-handside terms in
equations (22)-(23) become zero) are the origin together with the
entire $A$ axis (shown by a thick line). Each phase path corresponds
to a particular possible development of the system, indicated by an
arrow in Fig. 2, showing how the state of the system changes as time
increases. For example, if the system is initially in equilibrium
(i.e. only $A$ exists and $B$ is zero), then even a very small
displacement in $B$ leads to an evolution in the system: $B$
increases and $A$ decreases. When $B$ reaches to its maximum, $A$
becomes zero. Then the system continues development and reaches
again to the $A$ axis.

\begin{figure}
\includegraphics[width=8cm]{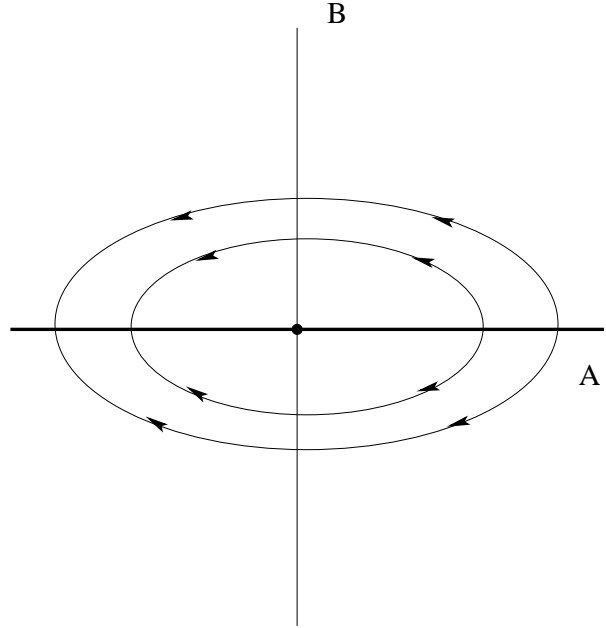}
\caption{\label{fig:epsart}Phase plane diagram for
$A=C_{10}/\rho_0$ (sound mode) and $B=C_{30}/B_0$ (Alfv{\'e}n mode) as described by equations
(22)-(23). Equilibrium points of the system are the origin and the
entire $A$ axis. Each phase path corresponds to a particular
possible development of the system, indicated by an arrow. It is
seen that even a very small displacement of the system from the
equilibrium points (i.e. from the $A$ axis) results in further
developments. Physically, it means that a small transversal
perturbation causes an energy transfer from the sound wave into the
Alfv{\'e}n wave. When $B$ reaches a maximum, the entire energy of
the system is stored in Alfv{\'e}n waves. Then the energy is once
more transferred into the sound wave. Note that if we draw the same
phase diagram for $\sqrt{(C^2_{30}+ C^2_{31})/B^2_0}=\sqrt{2}B$
(which is the real amplitude of Alfv{\'e}n wave), then the ellipse
transforms into a circle.}
\end{figure}

The potential energy density of a sound wave is
$E_s=C^2_{10}/\rho^2_0=A^2$ and the energy density of an Alfv{\'e}n
wave is $E_A=(C^2_{30} + C^2_{30})/B^2_0=2B^2$. Thus the phase
diagram shows that the waves continually exchange their energies: if
initially there is a sound wave, then the small transverse
displacement causes an energy transformation into an Alfv{\'e}n
wave, and {\it vice versa}. Equation (25) shows that the sum of wave
energies remains constant, as suggested on physical ground (in the
absence of dissipation).

We now turn to the long term evolution of sound $E_s$ and Alfv{\'e}n
$E_A$ wave energies, derived from a numerical solution of the
differential equation system (14)-(17). The results are plotted in
Figure 3. Only an Alfv{\'e}n wave exists initially. Subsequently,
its energy transforms into the sound wave before later returning
back to the Alfv{\'e}n wave, much as predicted analytically. In the
absence of dissipation, the waves exchange their energies
alternately.

\begin{figure}
\centering
\includegraphics[width=0.99\linewidth]{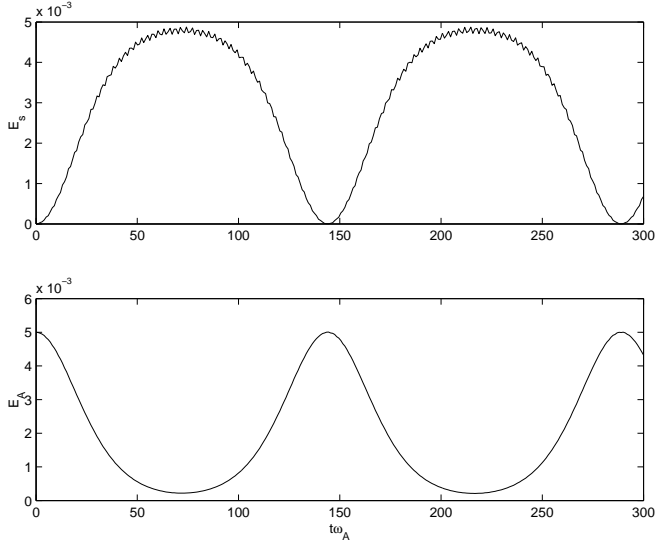}
\caption{The long term evolution of wave energy $E_s$ in the sound
mode and the wave energy $E_A$ in the Alfv{\'e}n mode. Initially the
energy is stored in the Alfv{\'e}n wave ($E_s = 0$). Subsequently it
transforms into the energy of a sound wave. An alternating energy
exchange between the waves then ensues. The process is averaged by
rapid oscillations.}
\end{figure}

The numerical solution of equations (14)-(17) shows that the time
scale of energy exchange depends on the wave amplitudes; it is
shorter for waves with stronger amplitudes. The dependence of
$\delta / \omega_A$ (the ratio of the energy transfer frequency
$\delta$ to the wave frequency $\omega_A$) on the wave amplitude,
derived by numerical integration of system (14)-(17), is displayed
in Figure 4.

Thus, as demonstrated both analytically and numerically, there is a
continuous energy exchange between sound and Alfv{\'e}n waves. The
energy transfer occurs in both ways: an initial sound wave transfers
its energy to an Alfv{\'e}n wave, and {\it vice versa}.

Now let consider the sound wave to be the pump, i.e. the amplitude
of the sound wave is much stronger than the initial amplitude of the
Alfv{\'e}n wave. Then the back reaction can be ignored with $C_1$
taken to be constant. Under these circumstances equation (18) takes
the form:
\begin{equation} {{d^2 C_3}\over dt^2} + 2i{\omega_A}{{d C_3}\over dt} +
{{\omega^2_A}\over {\rho_0}}C_1C^*_3 - {1\over {\rho_0}}C_1{{d^2
C^*_3}\over dt^2} = 0.
\end{equation}
We search for solutions of the form
\begin{equation}
C_1=C_{10}+iC_{11},
\end{equation}
\begin{equation}
C_3=(C_{30}+iC_{31})e^{\delta t},
\end{equation}
where $C_{10}$, $C_{11}$, $C_{30}$ and $C_{31}$ are constants.

Substituting expressions (27)-(28) into equation (26), and for
simplicity considering $C_{11}=0$, we obtain a fourth order equation
for $\delta$:
\begin{equation} \left ({1 - {C^2_{10}\over
{\rho^2_0}}}\right )\delta^4 + 2\omega^2_A\left ({2 + {C^2_{10}\over
{\rho^2_0}}}\right ) \delta^2 - {\omega^4_A}{C^2_{10}\over
{\rho^2_0}}=0.
\end{equation}
Then $C_3$ has an exponentially growing solution with the growth
rate
\begin{equation}
{\delta}{\approx}{1\over 2}{{C_{10}}\over {\rho_0}}{\omega_A}.
\end{equation}

\begin{figure}
\includegraphics[width=9cm]{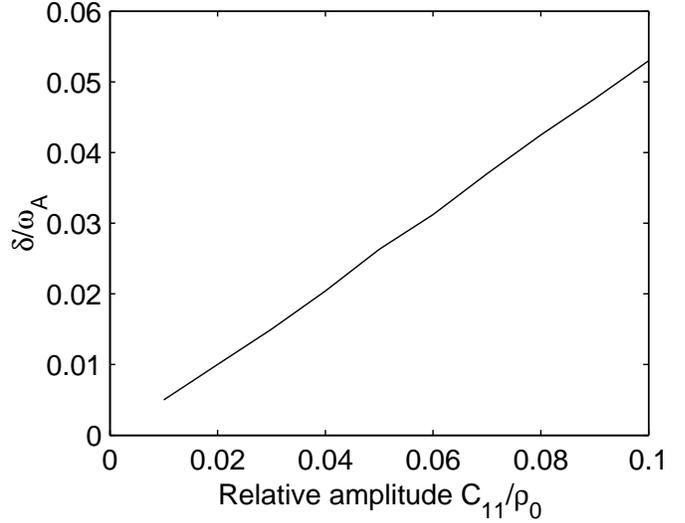}
\caption{\label{fig:epsart}The dependence of $\delta/ \omega_A$, the
ratio of the modulation frequency $\delta$ and wave frequency
$\omega_A$, on the wave amplitude derived by numerical calculations
of the system (14)-(17). The time scale of energy exchange is
shorter for strong amplitude waves.}
\end{figure}

Thus the pump sound wave causes an exponential amplification of the
Alfv{\'e}n wave amplitude. Energy transfers from the initial pump
sound wave to the Alfv{\'e}n wave. The reverse process, i.e. energy
transformation from a pump Alfv{\'e}n wave into sound waves, is
first studied by Hollweg (\cite{hol}).

Numerical solution of the system (14)-(17) when $C_1$ and $C_2$ are
constants also shows an exponential amplification of $C_3$ and
$C_4$, as suggested by the analytical solution (see equations (28)
and (30)). The results of a numerical solution, together with the
analytical expression (30), are plotted in Figure 5. It is seen that
the analytical and numerical solutions are indistinguishable (given
the same initial conditions).

Here we have solved numerically only the system (14)-(17), obtained
after averaging by rapid oscillation. In future it would be
interesting to investigate the complete numerical simulation of
equations (5)-(9).

\section{Discussion}

We have studied the weakly nonlinear interaction between sound and
linearly polarised Alfv{\'e}n waves propagating along an applied
magnetic field. We have shown that after averaging of the nonlinear
MHD equations for rapid oscillations the wave amplitudes become time
dependent if the frequencies and wave numbers satisfy the resonant
conditions $\omega_s={\pm}2\omega_A$ and $k_s={\pm}2k_A$, where
$\omega_A, k_A$ and $\omega_s, k_s$ are the frequencies and wave
numbers of Alfv{\'e}n and sound waves, respectively. These
conditions are fulfilled when the phase speeds of Alfv{\'e}n and
sound waves are equal, so that $v_A \approx v_s$. Then the waves
propagating in the same direction along the magnetic field
alternately exchange energy (see Figs. 2 and 3). The timescale of
energy exchange depends on the wave amplitude (see Fig. 4); strong
amplitude waves cause a faster process of energy exchange. Thus the
long standing uncertainty in the $\beta=1$ case in nonlinear MHD
may be resolved
by resonant nonlinear coupling between Alfv{\'e}n and sound waves.
However, generally astrophysical plasmas are highly
inhomogeneous, requiring a treatment of this case. 
In principle, a calculation along the lines described here for a 
homogeneous plasma can also be done in
the case of a transversally inhomogeneous medium.
Then the coupling may take place when the waves propagate in a
{\it resonant layer} where $v_A \approx v_s$.

\begin{figure}
\includegraphics[width=8cm]{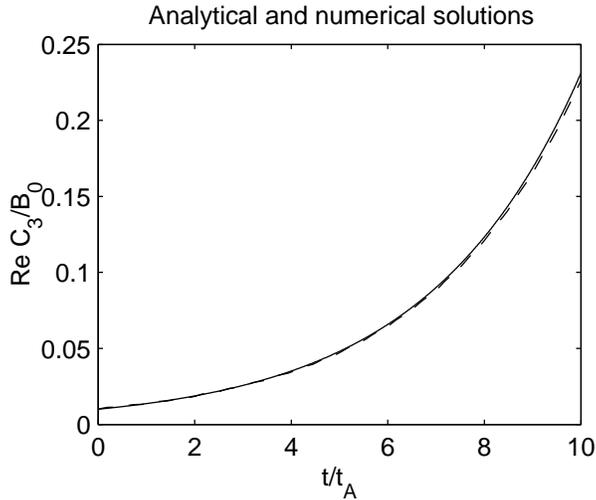}
\caption{\label{fig:epsart}The plots of analytical (solid line) and
numerical (dashed line) expressions of Alfv{\'e}n wave amplitude are
presented here when the sound wave is considered as the pumping
wave. The relative amplitude of sound wave $C_{10}/{\rho_0}$ is 0.1,
while initial relative amplitude of the Alfv{\'e}n wave $C_{30}/B_0$
is 0.01. The time is normalised by the period of Alfv{\'e}n wave,
$t_A$. The exponential growth of Alfv{\'e}n wave amplitude is
identical in both cases.}
\end{figure}

Note that here are only two coupled waves: sound and linearly
polarised Alfv{\'e}n waves propagating in the same direction along
the applied magnetic field. Therefore this process is basically
different from the well known three-wave interaction (Galeev and
Oraevsky \cite{gal}; Sagdeev \& Galeev \cite{sag}). We call it {\it
swing interaction} or {\it two-wave interaction}. Here the
Alfv{\'e}n wave drives the sound wave, due to the nonlinear magnetic
pressure, while the sound wave returns the energy back to the
Alfv{\'e}n wave through parametric action (Zaqarashvili \cite{zaq1};
Zaqarashvili \& Roberts \cite{zaq2}). The physics of the parametric
(swing) influence is simple: the sound wave causes a periodical
variation of the medium's plasma density, i.e. the fluid inertia,
and consequently works against the tension force (the restoring
force for an Alfv{\'e}n wave). The harmonics of an Alfv{\'e}n wave
with half the frequency of a sound wave grow in time, as in the case
of a pendulum with periodically varying length. The half frequency
arises due to the displacement of plasma elements on both sides of
the magnetic field (see Fig. 1).

Here we have shown that the wave coupling occurs in the region where
the waves propagate with the same speed, $v_A=v_s$, along the
magnetic field. However, it is well known that any resonant process
has a resonant interval of frequencies, the width of which depends
on the wave amplitudes. Therefore the sound and Alfv{\'e}n waves
with relatively stronger amplitudes may be coupled even in the case
when $v_A{\not =}v_s$. Thus the width of {\it resonant layer}
will be wider for stronger amplitude waves. Also, a similar
phenomenon may arise between Alfv{\'e}n and obliquely propagating
fast magnetosonic waves. But these processes require further study.

It must be mentioned that numerical simulations of wave propagation
in a two-dimensional stratified magneto-atmosphere show the coupling
between MHD waves in the region where the sound and Alfv{\'e}n
speeds are comparable in magnitude (Rosenthal et al. \cite{ros};
Bogdan et al. \cite{bog}). This very interesting result is different
from our consideration as the coupling in that paper occurs between
fast and slow magnetoacoustic-gravity waves. It would be interesting
to carry out numerical simulations in order to test the coupling
between sound and Alfv{\'e}n waves suggested in our paper.

It may be noted that the method of slowly varying amplitudes used
here does not include the process of wave steepening due to the
generation of higher harmonics. Consequently, it describes the
energy transformation process only in the early stages of wave
coupling. The generation of higher harmonics may modify the
situation. However, electron dispersion effects in two fluid MHD may
act against steepening and then the process of coupling will remain
unchanged.

Recent modelling of the plasma $\beta$ in the solar atmosphere (Gary
\cite{gary}) shows that $v_A \sim v_s$, i.e. $\beta \sim 1$, 
may takes place not
only in lower chromosphere, but also at relatively low coronal
heights. Thus the suggested wave coupling can be of importance
in that part of the solar atmosphere, where $v_A \sim v_s$, and in
the solar wind.

It is generally considered that solar 5-minute acoustic oscillations
($p$-modes) cannot penetrate into the corona due to the sharp
temperature gradient in the transition region. However, 5-minute
intensity oscillations are intensively observed in the corona by the
space satellites SOHO (Solar and Heliospheric Observatory) and TRACE
(Transition Region and Coronal Explorer) (De Moortel et al.
\cite{dem}), and recently De Pontieu et al. (\cite{dep1,dep2}) have
discussed how photospheric oscillations can be channelled into the
corona through inclined magnetic fields. Our proposed mechanism of
wave coupling may also resolve this problem. The acoustic
oscillations may transform their energy into Alfv{\'e}n waves, or
possibly into surface kink waves in thin photospheric magnetic
tubes, this process acting in the region of the solar atmosphere
where $v_A{\approx}v_s$. Generated transversal waves may then
propagate through the transition region into the corona, where they
can deposit their energy back into density perturbations. The
process can be thus be of importance in coronal heating. It is
interesting to note that MHD oscillations with the properties of
Alfv{\'e}n waves have been observed in the photosphere and lower
chromosphere by Ulrich (\cite{ulr}). He found that the power
spectrum of the magnetic variations includes substantial power at
frequencies lower than the 5 minute oscillation (see Fig. 3 of that
paper). This may be caused by our energy conversion mechanism. It is
also interesting to note the recent observations by Muglach et al.
(\cite{mug}), which suggest a possible transformation of
compressible wave energy into incompressible waves in the $\beta
\approx 1$ region of the solar atmosphere.

Finally, we note that the two-wave interaction process can also be
of importance as an explanation of the observed rotational
discontinuities and pressure-balanced structures found in the solar
wind (Vasquez \& Hollweg \cite{vas1,vas2}). Swing coupling between
Alfv{\'e}n and obliquely propagating fast magnetosonic waves may
lead to the formation of these structures, and this in turn can be
of importance in the problem of solar wind acceleration.

\section{Conclusions}

We have described the nonlinear coupling between sound and linearly
polarised Alfv{\'e}n waves propagating with the same speed along an
applied magnetic field. The sound wave is coupled to the Alfv{\'e}n
wave with a period and wavelength that is double that of the
Alfv{\'e}n wave. Analytical and numerical solutions show that the
waves alternately exchange their energies during propagation. 
The phenomenon also can be of
importance (after appropriate modifications) in the solar
atmosphere and solar wind as well as in
various other astrophysical and laboratory situations.

\begin{acknowledgements}
We thank Professors J. V. Hollweg (UNH) and M. S. Ruderman
(Sheffield) for helpful discussions. The work of T. Z. was partially
supported by the NATO Reintegration Grant FEL.RIG 980755 and MCyT
grant AYA2003-00123.
\end{acknowledgements}

{}

\end{document}